\documentclass[twocolumn,superscriptaddress,showpacs,preprintnumbers,amsmath,amssymb]{revtex4}
\usepackage{graphicx}% Include figure files
\usepackage{dcolumn}% Align table columns on decimal point
\usepackage{bm}% bold math

\begin{document}

\preprint {}

\title{Writing and Reading antiferromagnetic Mn$_2$Au: N{\'e}el spin-orbit torques and large anisotropic magnetoresistance.}% Force line breaks with \\

\author{S. Yu. Bodnar}
\affiliation{Institut f\"ur Physik, Johannes Gutenberg-Universit\"at, Staudinger Weg 7, 55128 Mainz, Germany}
\author{L. \v{S}mejkal}
\affiliation{Institut f\"ur Physik, Johannes Gutenberg-Universit\"at, Staudinger Weg 7, 55128 Mainz, Germany}
\affiliation{Institute of Physics, Academy of Sciences of the Czech Republic, Cukrovarnicka 10, 162 00 Praha 6, Czech Republic}
\affiliation{Faculty of Mathematics and Physics, Charles University, Department of Condensed Matter Physics, Ke Karlovu 5,
12116 Praha 2, Czech Republic}
\author{I. Turek}
\affiliation{Faculty of Mathematics and Physics, Charles University, Department of Condensed Matter Physics, Ke Karlovu 5,
12116 Praha 2, Czech Republic}
\author{T. Jungwirth}
\affiliation{Institute of Physics, Academy of Sciences of the Czech Republic, Cukrovarnicka 10, 162 00 Praha 6, Czech Republic}
\affiliation{School of Physics and Astronomy, University of Nottingham, University Park, Nottingham NG7 2RD, United Kingdom}
\author{O. Gomonay}
\affiliation{Institut f\"ur Physik, Johannes Gutenberg-Universit\"at, Staudinger Weg 7, 55128 Mainz, Germany}
\author{J. Sinova}
\affiliation{Institut f\"ur Physik, Johannes Gutenberg-Universit\"at, Staudinger Weg 7, 55128 Mainz, Germany}
\author{A.A. Sapozhnik}
\affiliation{Institut f\"ur Physik, Johannes Gutenberg-Universit\"at, Staudinger Weg 7, 55128 Mainz, Germany}
\author{H.-J. Elmers}
\affiliation{Institut f\"ur Physik, Johannes Gutenberg-Universit\"at, Staudinger Weg 7, 55128 Mainz, Germany}
\author{M. Kl\"aui}
\affiliation{Institut f\"ur Physik, Johannes Gutenberg-Universit\"at, Staudinger Weg 7, 55128 Mainz, Germany}
\author{M. Jourdan}
 \email{Jourdan@uni-mainz.de}
\affiliation{Institut f\"ur Physik, Johannes Gutenberg-Universit\"at, Staudinger Weg 7, 55128 Mainz, Germany}

\date{\today}% It is always \today, today,

%\begin{abstract}

%\end{abstract}

%\pacs{needs input}% PACS, the Physics and Astronomy
%\keywords{}%Use showkeys class option if keyword
\maketitle

{\bf Antiferromagnets are magnetically ordered materials which exhibit no net moment and thus are insensitive to magnetic fields. Antiferromagnetic spintronics \cite{Jun16} aims to take advantage of this insensitivity for enhanced stability, while at the same time active manipulation up to the natural THz dynamic speeds of antiferromagnets \cite{Kam11} is possible, thus combining exceptional storage density and ultra-fast switching. However, the active manipulation and read-out of the N{\'e}el vector (staggered moment) orientation is challenging. Recent predictions have opened up a path based on a new spin-orbit torque \cite{Zel14}, which couples directly to the N\'{e}el order parameter. This N\'{e}el spin-orbit torque was first experimentally demonstrated in a pioneering work using semimetallic CuMnAs \cite{Wad16}. Here we demonstrate for Mn$_2$Au, a good conductor with a high ordering temperature suitable for applications, reliable and reproducible switching using current pulses and read-out by magnetoresistance measurements. The symmetry of the torques agrees with theoretical predictions and a large read-out magnetoresistance effect of more than ${\bf \simeq 6}$~${\bf \%}$ is reproduced by {\em ab initio} transport calculations.}

For the key application operations of reading and writing in antiferromagnets, different approaches have been previously put forward. Initial experiments on spin-valve structures with an antiferromagnet (AFM) as the active layer manipulated the N{\'e}el vector by an exchange-spring effect with a ferromagnet (FM) and read-out via Tunneling-Anisotropic Magnetoresistance (T-AMR) measurements \cite{Par11}. Other related experiments were based on the same effect \cite{Fin14}, or on a FM to AFM phase transition \cite{Mar14}. However, the most promising approach is to use current induced spin-orbit torques for switching the N{\'e}el vector. It exhibits superior scaling and its counterpart in ferromagnets is already established and considered among the most efficient switching mechanisms for memory applications. \cite{Gam11, Bra14}.

Only two compounds, CuMnAs and Mn$_2$Au, are known to provide at room temperature the collinear commensurate antiferromagnetic order and specific crystal structure, which is predicted to result in the staggered spin accumulation in the sublattice structure, leading to bulk N{\'e}el spin-orbit torques allowing for current induced switching of the N{\'e}el vector \cite{Zel14}.

Semimetallic CuMnAs was grown previously by molecular beam epitaxy (MBE) with a N{\'e}el temperature of $\simeq 500$~K \cite{Wad13} and current-induced switching of these samples was recently demonstrated for the first time \cite{Wad16, Grz17}. However, for spintronics applications the compound Mn$_2$Au provides several advantages, as it is a good metallic conductor and does not contain toxic components. Furthermore, its magnetic ordering temperature is well above 1000~K \cite{Bar13}, providing the necessary thermal stability for applications. Mn$_2$Au shows a simple antiferromagnetic structure with the collinear magnetic moments in the (001)-plane \cite{Shi10, Bar13, Bar16}. Thin film samples were previously grown in (101)-orientation by MBE \cite{Wu12} and Fe/Mn$_2$Au(101) bilayers showed AMR effects of up $2.5$~$\%$ in a 14~T rotating magnetic field \cite{Wu16}.

While Mn$_2$Au was the first compound for which current-induced internal staggered spin-orbit torques were predicted \cite{Zel14}, corresponding experimental evidence has been missing. Here we report current induced N{\'e}el vector switching in Mn$_2$Au(001) epitaxial thin films, which is is easily read-out by a large AMR. 

Our Al$_2$O$_3$/Ta(10 nm)/Mn$_2$Au(75 nm)/Ta(3 nm) samples were prepared by sputtering as described elsewhere \cite{Jou15} and patterned into a star-structure as shown in Fig.\,1.
\begin{figure}[htb]
\includegraphics[width=0.8\linewidth]{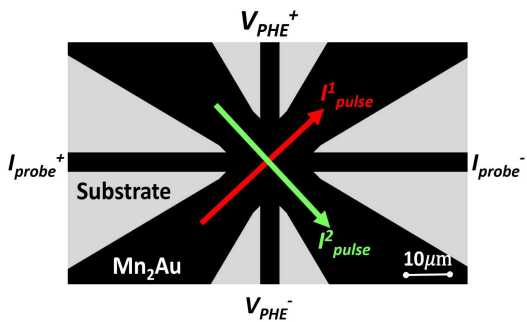}
\caption{Sample layout (star pattern) used for the current induced N{\'e}el vector manipulation experiments with current pulse directions $I_{pulse}^{1/2}$ and probing contacts for PHE and AMR measurements indicated.}
\end{figure}
This geometry allows for electric writing of the N{\'e}el vector orientation by pulsing currents along the two perpendicular directions $I_{pulse}^1$ and $I_{pulse}^2$ and for electric read-out by measuring either the transversal resistivity $\rho_{xy}$, i.\,e.\, the Planar Hall Effect (PHE), or the longitudinal resistivity $\rho_{long}$, corresponding to the AMR of the samples. Depending on the orientation of the patterned structure, the pulse currents can be sent along different crystallographic directions.  

The AMR of a single domain sample is given by
\begin{equation}
 {\rm AMR_{hkl}}=\frac{\rho_{long}(\phi=0^{\rm o})-\rho_{long}(\phi=90^{\rm o})}{\bar{\rho}_{long}}=\frac{\Delta\rho_{long}}{\bar{\rho}_{long}},
\end{equation}
where $\rho_{long}$ is longitudinal resistivity, $\phi$ is the angle between the N{\'e}el vector and current direction and [hkl] is the N{\'e}el vector orientation in the basis of the tetragonal conventional unit cell.
The PHE usually observed in ferromagnetic materials scales with the AMR and shows a dependence on the angle $\phi$ given by \cite{Tom75,See11}:
\begin{equation}
\rho_{xy}=\Delta \rho_{long}\, \sin\phi \cos\phi
\end{equation}
Thus also in antiferromagnets $\rho_{xy}$ has its maximum value and changes sign if $\phi$ switches from $+45^{\rm o}$ to $-45^{\rm o}$.

To study the switching, trains of $100$ current pulses with a pulse length of $1$~ms and a delay between the pulses of 10~ms were applied. As after a pulse train thermal relaxation behaviour on a time scale of $1$~s after was observed, the read-out was performed with a delay of $10$~s.  
Fig.\,2 shows the transversal resistivity $\rho_{xy}$ versus the number of applied pulse trains.
\begin{figure}[htb]
\includegraphics[width=1.1\linewidth]{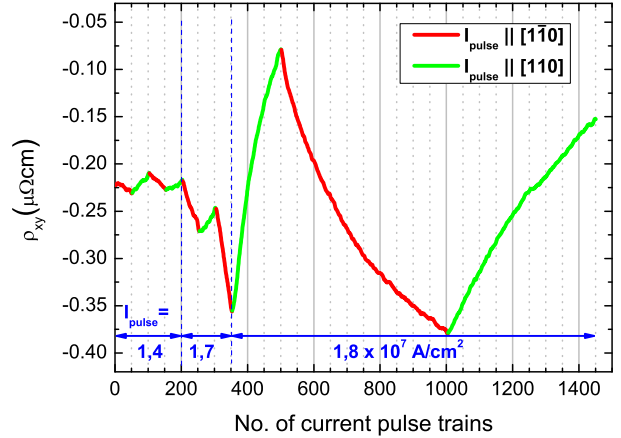}
\caption{Probed transversal resistivity (DC probing current density $10^4$~A/cm$^2$) vs. number of applied pulse trains along different directions. The crystallographic direction in which the current pulses were injected is indicated by the green and red color of the data points. The pulse current density was increased from $1.4\times10^7$~A/cm$^2$ to $1.8\times10^7$~A/cm$^2$ as indicated in the graph.}
\end{figure}
First, a pulse current density of $1.4\times10^7$~A/cm$^2$ was applied along the [1$\bar{1}$0] direction, resulting in a small change of the corresponding Hall voltage. Without reaching saturation after $50$ pulse trains the pulse current direction was switched to [110], resulting in a reversal of the corresponding change of the transversal resistivity. This sequence could be reproduced several times. Increased pulse current densities of $1.7\times10^7$~A/cm$^2$ and $1.8\times10^7$~A/cm$^2$ resulted in larger changes of the corresponding Hall voltages. By increasing the number of pulse trains applied along the [110] direction to $500$, a trend towards saturation of the Hall voltage was obtained.

Internal field like spin-orbit torques are expected to generate reversible switching between distinct stable states if the current is injected along biaxial easy directions \cite{Zel14,Roy16}. However, we observed reversible switching to stable states for pulse currents along both the crystallographic [110] and [100] axes (rotated star-pattern). Thus we conclude that the in-plane magnetic anisotropy of our Mn$_2$Au thin films is weak. This is consistent with our calculations of the magnetocrystalline anisotropy energy (MAE), which is almost negligible within the ab-plane (see {\em Methods}).

An example of the resulting changes of the transversal and longitudinal resistivities generated by pulse currents along the [100] and [010] directions is displayed in Fig.\,3.
\begin{figure}[htb]
\includegraphics[width=1.0\linewidth]{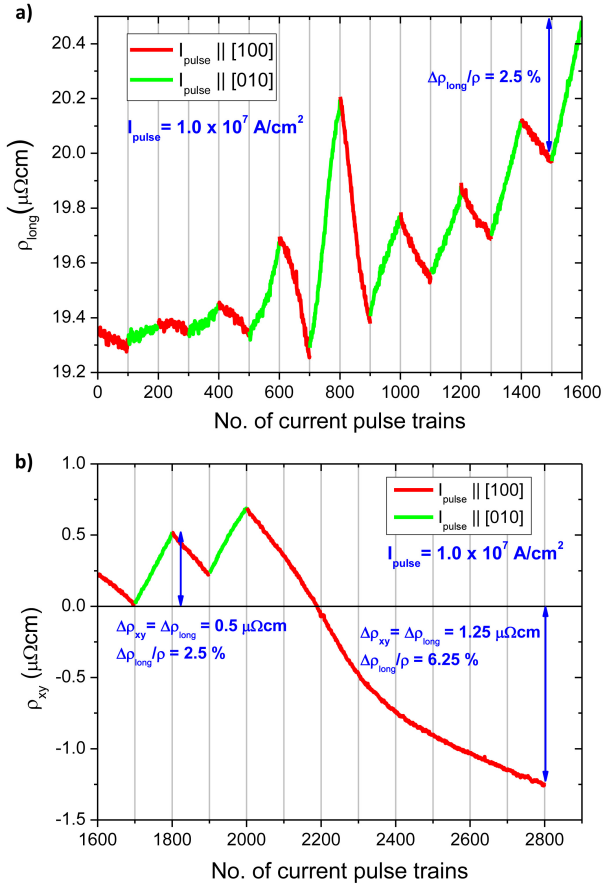}
\caption{(a) Longitudinal resistivity (DC probing current density $10^4$~A/cm$^2$) vs.\,number of applied pulse trains. 
(b) Transversal resistivity of the same sample vs.\,number of applied pulse trains. The crystallographic direction in which the current pulses were injected is indicated by the green and red color of the data points.}
\end{figure}
The upper panel (Fig.\,3a) shows the longitudinal resistivity probed after each of the first 1600 pulse trains consisting of 100 pulses each with a current density $1.0\times10^7$~A/cm$^2$. For the first sequences only small variations of the longitudinal resistivity were observed. However, with the application of subsequent pulse trains the magnitude of the effect increased. This training-like behaviour may be associated with the motion and pinning of AFM domain walls in the sample. After 1600 pulse trains a constant resistance change of $\Delta\rho_{long} / {\bar \rho} = 2.5$~\% induced by 100 pulse trains was reached, which is an order of magnitude higher than what is observed for CuMnAs \cite{Wad16}.

To check the origin of these changes, the transversal resistivity of the sample was measured and also showed reproducible pulse current induced changes (Fig.\,3b). The increase of the transversal resistivity induced by 100 pulse trains amounted to $\rho_{xy}=0.5$~${\rm \mu\Omega cm}$.

Based on these numbers the identification of the longitudinal and transversal resistivities with the AMR and PHE can be verified: If both effects originate from the same anisotropic electron scattering, they have to be related by equation (2). We assume a switching of the N{\'e}el-vector in parts of the sample corresponding to a change of $\phi$ in equation (2) from  $+45$~$^{\rm o}$  to  $-45$~$^{\rm o}$, i.\,e.
\begin{equation}
 \Delta\rho_{xy}=\rho_{xy}(+45^{\rm o})-\rho_{xy}(-45^{\rm o})=\Delta\rho_{long},
\end{equation}
Thus we find that $\rho_{xy}=0.5$~${\rm \mu\Omega cm}$ corresponds again to $\Delta\rho_{long} / {\bar \rho} = 2.5$~\%. This consistency of the longitudinal and transversal resistivities provides strong evidence for an intrinsic electronic origin of the pulse current induced changes of the magnetoresistance signals.

After two more pulse current direction reversals reproducing the previous behaviour of the sample, the pulse current direction was kept along [100] for 800 additional pulse trains. This resulted in a sign reversal of the PHE. Although a small offset of the transversal voltage due to e.\,g.\,imperfections of the patterned structure is possible, the magnitude of the PHE indicates that the N{\'e}el vector now switched in the majority of the sample.   After about 300 pulse trains along the [100] direction a beginning saturation of the PHE resistivity appeared, but was not completed when after 500 additional pulse trains the sample broke. A maximum transversal resistivity of $\rho_{xy}=1.25$~${\rm \mu\Omega cm}$ was reached, which based on equation (3) corresponds to an AMR of $6.25$~$\%$. This is one of the largest found in metallic magnetic thin films, and its size bodes well for easy read-out of the antiferromagnetic state as necessary for device applications. While small variations exist between samples, we observe consistently larger AMR effects for pulse currents along the [100] than for the [110] directions.

To understand the origin of the magnetoresistance effects, we calculated the AMR of single domain Mn$_2$Au, assuming a complete $90^{\rm o}$ switching of the N{\'e}el vector (see {\em Methods}). In general, AMR originates from the effects of spin-orbit coupling on the band structure and from scattering from an extrinsic disorder potential \cite{Ran08}. When incorporating the effects of realistic disorder in the calculations (see {\em Methods}), two types were considered: Off-stoichiometry and inter-site swapping between Mn and Au atoms. Experimentally, the former was analyzed by energy dispersive x-ray spectroscopy (EDX) of $500$~nm thick Mn$_2$Au films resulting in a stoichiometry of $66.2\pm 0.3$~$\%$ Mn and $33.8\pm 0.3$~$\%$ of Au, which indicates a slight Au excess. Also a small degree of inter-site disorder can be expected, but its quantification is experimentally not accessible. Thus we simulated a slight excess of Au randomly distributed over the Mn sites and random Mn - Au swapping. 

We calculated the AMR for two crystal directions of the N{\'e}el vector, AMR$_{100}$ and AMR$_{110}$. Fig.\,4a, shows the results for different degrees of disorder in Mn$_2$Au.
\begin{figure}[htb]
\includegraphics[width=1.0\linewidth]{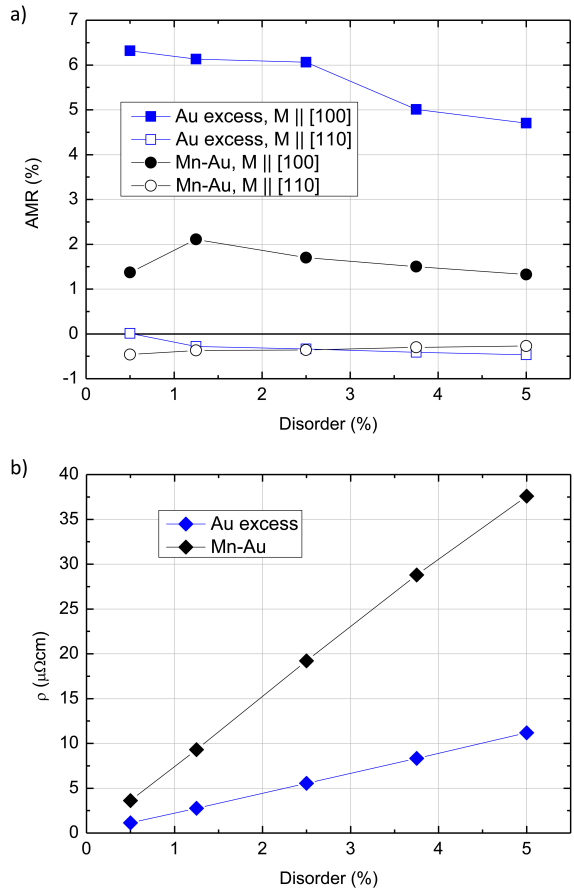}
\caption{(a) Calculated AMR of Mn$_{2}$Au for different degrees of disorder due to Au excess and due to Mn - Au site swapping with dependence on the N{\'e}el vector orientation.
(b) Calculated residual resistivities of Mn$_{2}$Au for different degrees disorder due to Au excess and due to Mn - Au site swapping.}
\end{figure}
Large AMR values consistent with our experiments were obtained for small degrees of disorder reaching a maximum value of $6.3$~\% for 0.5~\% excess of Au. Moreover, we obtain $AMR_{100}>AMR_{110}$, which reproduces the experimentally observed trend for the two crystalline directions. The corresponding calculated residual resistivities as shown in Fig.\,4b are consistent with the experimentally obtained values ($\simeq 8$~${\rm \mu\Omega cm}$ \cite{Jou15}), corroborating the relevance of the simulated type of disorder.
 
In summary, in-plane switching of the N{\'e}el vector in the antiferromagnetic metal Mn$_2$Au by current pulses was realized using intrinsic spin-orbit torques. Consistent measurements of the AMR and PHE showed pulse current direction dependent reversible changes, providing direct evidence for N{\'e}el vector switching. Easy read-out of the switching is provided by a large amplitude of the AMR of more than 6\%, which is more than an order of magnitude higher than previously observed for other antiferromagnetic systems and one of the highest AMR amplitudes found for metallic magnetic thin films. We can reproduce the magnitude of the effect theoretically by including realistic disorder and, in particular, find the same dependence of the amplitude on the crystallographic directions in the experiment as in the calculation. With the basic principles of writing and read-out demonstrated, combined with a theoretical understanding of the underlying spin-orbit torques, and the large magnetoresistive effects, the metallic compound Mn$_2$Au is a prime candidate to enable future AFM spintronics.

{\bf Acknowledgments:} This work is supported by the German Research Foundation (DFG) through the Transregional Collaborative Research Center SFB/TRR173 {\em Spin+X}, Projects A03 and A05. J.S., L.S., O.G., and T.J. acknowledge the support of the Alexander von Humboldt Foundation, the ERC Synergy Grant SC2 (No.\,610115), the Ministry of Education of the Czech Republic Grant No. LM2015087, and the Grant Agency of the Czech Republic grant no. 14-37427G, L.S. acknowledges support from the Grant Agency of the Charles University, no.\,280815.\,Access to computing and storage facilities owned by parties and projects contributing to the National Grid Infrastructure MetaCentrum provided under the program "Projects of Large Research, Development, and Innovations Infrastructures" (CESNET LM2015042), is greatly appreciated. The work of I.T. was supported by the Czech Science Foundation (Grant No.\,14-37427G).

{\bf Author contributions}

S.Yu.B. and M.J. mainly wrote the paper and performed the transport measurements; L.S. performed the AMR and anisotropy calculations and wrote the corresponding part of the manuscript, I.T. developed the codes for the transport calculations, S.Yu.B. and A.A.S. prepared the samples, H.-J.E, M.K., O.G., T.J., I.T, and J.S. discussed the results and contributed to the writing of the manuscript; M.J. coordinated the project.

{\bf Methods}

{\footnotesize
{\bf Measurement Procedure}
The current pulse trains were generated by a Keithley 2430 Source Meter. After each pulse train a delay of $10$~s for thermal relaxation was followed by a measurement of the transversal or longitudinal voltage across the central part of the patterned structure resulting from a probe current density of $10^4$~A/cm$^2$. Typically this procedure was repeated several times before the pulse trains were sent along the perpendicular direction of the cross structure, keeping the probing contacts unchanged.   

Based on time resolved resistivity measurements during the application of pulse trains and temperature dependent resistivity measurements of Mn$_2$Au thin films the local temperature of the relevant sample region was estimated: Values of up to $600$~K in the stable regime and  $\simeq 800$~K at current densities, which finally destroyed the samples, were obtained. The local temperature rapidly decays after the application of a current pulse train, therefore thermoelectric voltages can be neglected. 

{\bf MAE and AMR calculation}
The MAE was calculated using the FLAPW (Full Potential Linearized Augmented Plane Wave) method in combination with the GGA (Generalized Gradiend Approximation). We found the MAE in line with previous reports \cite{Shi00} $\simeq 1$~${\rm \mu}$eV per formula unit, which is at the resolution limit of our method based on the magnetic force theorem.

To calculate the AMR in Mn$_2$Au ab initio we employed the fully relativistic Dirac tight binding-linear muffin-tin orbital plus coherent potential approximation (FRD-TB-LMTO+CPA) method in combination with the Kubo formula \cite{Tur02, Tur14}. We used the s-, p-, and d-type orbitals in the basis and the LSDA (Local Spin Density Approximation) Vosko-Wilk-Nusair exchange-correlation potential parametrization \cite{Vos80}. The ground-state magnetization and density of states was reproduced consistently with a previous reports \cite{Shi10, Zel14}. In the transport calculation we used up to $10^8$ k-points in the Brillouin zone and for the residual resistivity calculations we set the imaginary part of the complex energy to $\Gamma=10^{-5}$~Ry.
}

\end{document}